\documentstyle[aps,multicol,prl,epsf]{revtex}

\begin{document}
\title{Theory of the Resistive Transition in Overdoped
$Tl_2Ba_2CuO_{6+\delta}$: 
Implications for the angular dependence of the quasiparticle scattering rate
in High-$T_c$ superconductors.}
\author{V. B. Geshkenbein$^{1,4}$, L. B. Ioffe$^{2,4}$ and A. J. Millis$^3$}
\address{$^1$Theoretische Physik, ETH H\"onggerberg, Z\"urich, Switzerland \\
$^2$ Department of Physics and Astronomy Rutgers University, Piscataway, NJ
08854 \\
$^3$ Department of Physics and Astronomy, The Johns Hopkins University\\ 3400
North Charles St., Baltimore, MD 21218 \\
$^4$ Landau Institute for Theoretical Physics, Moscow, Russia}

\maketitle

\begin{abstract}
We show that recent measurements of the magnetic field dependence of the
magnetization, specific heat and resistivity of 
overdoped $T_c \sim 17K$ $Tl_{2}Ba_{2}CuO_{6+\delta }$
in the vicinity of the superconducting $H_{c2}$ imply that the vortex
viscosity is anomalously small and that the material studied is
inhomogeneous with small, a few hundred $\AA$, regions in which the local
$T_{c}$ 
is much higher than the bulk $T_{c}$. The anomalously small vortex viscosity
can be derived from a microscopic model in which the quasiparticle lifetime
varies dramatically around the Fermi surface, being small everywhere {\it %
except} along the zone diagonal ({\it ``cold spot''}).
We propose experimental tests of our results.
\end{abstract}


\begin{multicols}{2}

The nature of the superconducting transition in high-$T_{c}$ superconductors
is the subject of present controversy: some workers argue that the
transition is driven by thermal or quantal fluctuations of the
superconducting order parameter \cite{Doniach90,Emery95} implying that in a
wide regime above the resistively determined  $T_{c}$ one
has local superconducting order without global phase coherence. In such a
situation one would  expect strong superconducting fluctuations,
i.e.  a large paraconductivity,
magnetoresistance and fluctuation diamagnetism. The apparent absence (or
weakness) of these effects in all but the most heavily underdoped compounds
casts doubt on the phase fluctuation hypothesis \cite{Geshkenbein97}.

Recent measurements  on  overdoped $T_{c0}= 17K$ $%
Tl_{2}Ba_{2}CuO_{6+\delta }$ in a magnetic field
provide an interesting new perspective on this
issue.  There is a reasonably sharp
resistive transition with onset temperature $T_{\rho }(H)$
\cite{MacKenzie93,Carrington94,Tyler97}. A
pairing onset temperature  $T_{\gamma }(H)$ 
can be defined for $H<4\;\mbox{T}$ from specific heat
and magnetization measurements \cite{Carrington96,Bergemann97}. At
$H=0$, $T_{\rho}=T_{\gamma}=T_{c0}$, the zero field transition temperature,
but at $H>0$ it is found
that $T_{\rho }(H)<T_{\gamma }(H)$, implying
that the resistive transition is due to vortex lattice melting.
However, in the range $T_{\rho }(H)<T<T_{\gamma }(H)$ the
reported resistivity has a very weak temperature and field dependence,
i.e. it
is possible in this material 
to destroy superconductivity by phase fluctuations without
producing a strong paraconductivity or magnetoresistance. The goal
of this paper is to understand in more detail this interesting phenomenon.

The $Tl_{2}Ba_{2}CuO_{6+\delta }$ materials have also attracted attention
because the resistively determined upper critical field $H_{\rho }(T)$ has
an anomalous temperature dependence \cite{MacKenzie93}, curving sharply
upwards as T is decreased below $5K$. 
It has been  widely assumed that the upward curvature is
an intrinsic property of the material \cite{alt.theories}, 
however we will argue that the low-T
behavior is due to the presence in the sample of small regions
with $T_c$ much higher than the bulk $T_{c0}$. A closely related idea
involving small regions with anomalously large $H_{c2}$ was put forward in the
general context of dirty superconductors in \cite{Spivak95}.

We first discuss the resistivity, specific heat and magnetization data. At low
applied magnetic field $B$ the material is superconducting.
The resistivity $\rho =0$ and there is a large 'London' magnetization given
in equilibrium by $M_{London}=-\Phi _{0}\ln (H_{c2}/B)/32\pi ^{2}\lambda ^{2}$%
.(Real materials are not in equilibrium because the vortex lattice is pinned).
 At some field $H_{melt}(T)$ (visible e.g. as
the foot of the resistive transition in Fig 2 of ref
\cite{MacKenzie93})  the
vortex lattice (glass) melts. 
The magnetization is observed to take the London value \cite{Bergemann97}
and the resistivity becomes non-zero.
For $H_{melt}(T)<H<H_{\rho}(T)$ the
resistivity has a strong $H$ and $T$ dependence; we interpret this as
evidence that the transport is dominated by vortex pinning near the melting
transition.  Above
$H_{\rho }(T)$ the resistivity saturates and loses its strong
field dependence. For concreteness we define $H_{\rho }(T)$ as the field at
which $\rho $ reaches $90\%$ of its saturated value. 

The specific heat exhibits a maximum which broadens and
shifts to lower $T$ as the field is increased. We define $H_{\gamma }(T)$ from 
the temperature of the maximum; it turns out that $H_{\gamma }(T)\approx
2H_{\rho }(T)$ in the field range $0<H<2\;\mbox{T}$ in which the maximum is
visible. 
We interpret $H_{\gamma }(T)$ as the scale at which bulk superconducting
pairing vanishes, i.e. as the 'microscopic' $H_{c2}$.

A paired state should exhibit a large magnetization vanishing at the
'microscopic' $H_{c2}$.  The observed \cite{Bergemann97}
magnetization is apparently given by the sum of a 'London' term with a $\ln
(1/B)$ field dependence and a diamagnetic term with a linear B dependence,
i.e. $M(B,T)=M_{London}+\chi _{d}B$.
The London term is observed for a range of temperatures greater than 
$T_{\rho}(H)$, however as $T$ approaches $T_{\gamma}(H)$ the second,
diamagnetic term dominates the measured magnetization so the behavior of the
London term is difficult to determine. The data are consistent
with the expectation that the London term vanishes at
$H_{\gamma}(T)$. The coefficient of the diamagnetic
term, $\chi_d$ has magnitude much larger than the usual Landau diamagnetism
and persists over a wide range of $T>T_{\gamma}(H)$, indeed up to $T \sim
2T_{co}$. We interpret the diamagnetic term as arising from the presence in
the sample of small regions with transition temperatures much higher than
the bulk superconducting $T_{co}$.

The scales $H_{melt}(T)$, $H_{\rho }(T)$ and $H_{\gamma }(T)$ are shown in
Fig 1, along with a shaded area indicating the region in the $H-T$ plane
where an anomalously large diamagnetic term $\chi _{d}$ is observed. Note
that a straight-line extrapolation of $H_{\gamma }(T)$ to $T=0$ yields a $%
H_{\gamma }(0)\ll H_{\rho }(0)$. The relatively small value of $H_{\gamma
}(T=0)$ implies that the low-T upturn of $H_{\rho }$ is not a bulk property
of the material. We will show below that it is due to the same 
inhomogeneities which
produce the diamagnetism.

\vspace{0.25cm}

\centerline{\epsfxsize=10cm \epsfbox{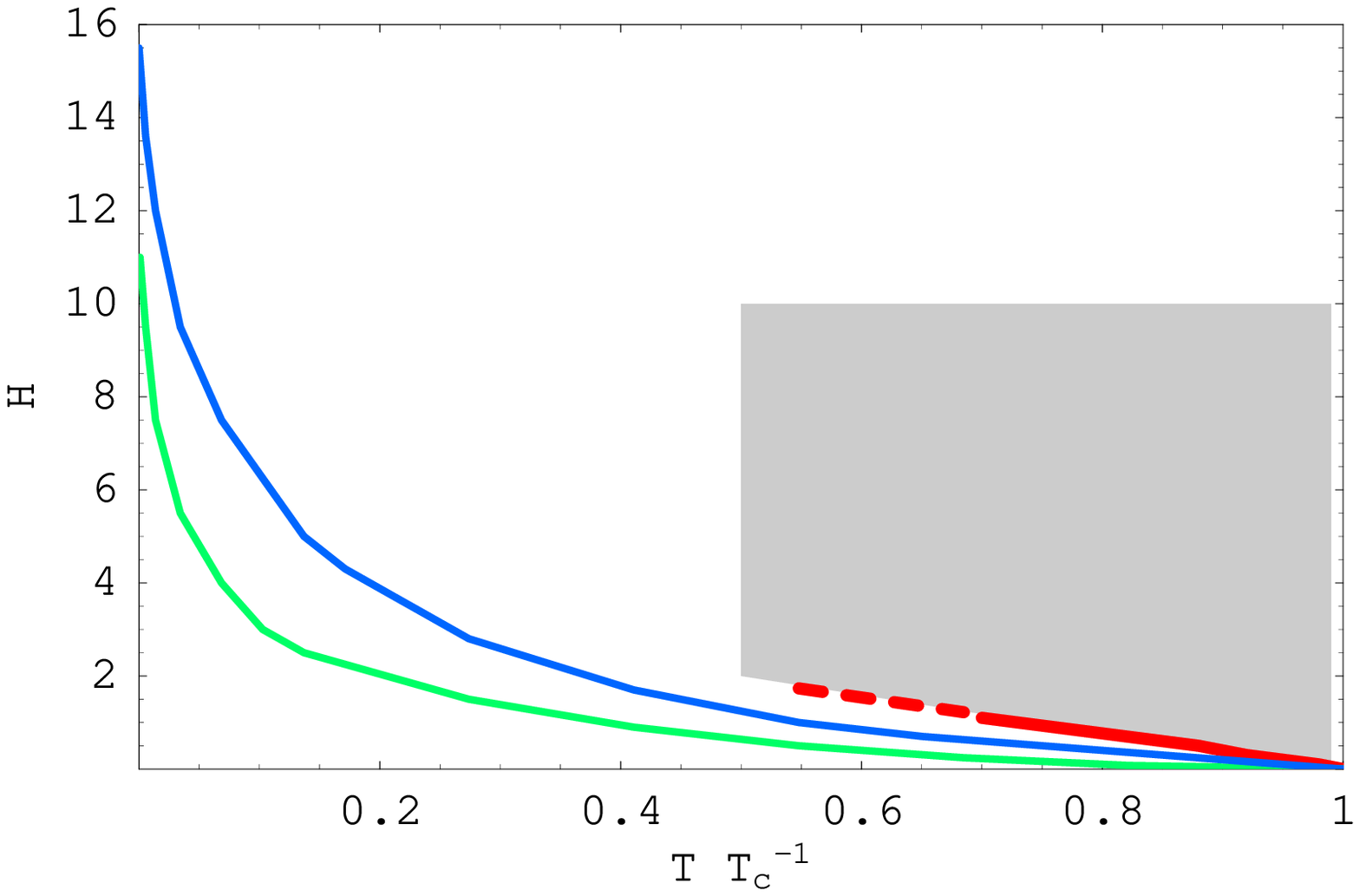}}

{\footnotesize
{\bf Fig 1} Phase diagram of $T_c \approx 15\;K$   $Tl_2Ba_2CuO_{6+\delta}$ as
determined from resistivity, specific heat and magnetization measurements.
Lowest line:  $H_{melt}$; intermediate line $H_{\rho}$,
upper line $H_{\gamma}$. Linear diamagnetism was observed in the shaded region
and (although not shown in the figure) persists to $T\approx 2T_c$.
}

\vspace{0.25cm}

From the phase diagram one sees a wide separation
between the local pairing scale $H_{\gamma }$ and the
resistive scales $H_{\rho }$ and $H_{melt}$. This implies that
the whole temperature range between $T_{melt}(H)$ and $T_{\gamma }(H)$
should be described as a vortex liquid, albeit one with unconventional 
transport
properties. The conventional view \cite{Tinkham} of transport in
a type II superconductor in a non-zero magnetic field is as follows. One has
vortices, these move in response to an applied current and this motion
causes the phase of the superconducting order parameter to become time
dependent leading via the Josephson relation to an electric field which causes
dissipation and hence a finite conductivity, $\sigma _{V}$. The total
conductivity, $\sigma $, is the sum of the vortex part, $\sigma _{V}$, and a
normal one, $\sigma _{n}$ due to uncondensed carriers. 
The standard estimate of the vortex conductivity in the flux-flow regime
is the Bardeen-Stephen formula 
\begin{equation}
\sigma _{BS}=\frac{H_{c2}}{H}\sigma _{n}.  \label{sigma_BS}
\end{equation}
Although it was originally obtained from phenomenological arguments,
subsequent work \cite{Kopnin95,Kopnin97} based on the quasiclassical kinetic
equation for superconductors shows that it is remarkably robust,
and applies in almost all situations except very close to the microscopic $%
H_{c2}$ or in the ''superclean'' limit.

The interpretation of the Bardeen-Stephen formula is that the density of
vortices is $H/\Phi _{0}$ ($\Phi _{0}$ is flux quantum) and the
dissipation per vortex
is the core area (which is $\Phi _{0}/H_{c2}$) divided by $\sigma
_{n}$. At $H\ll H_{c2}$ the number of vortices is small and the vortex
conductivity is large, $\sigma _{BS}\gg \sigma _{n}$. The observed
conductivity is then controlled by the vortices which short circuit any
conductivity from the normal carriers. It has a $1/H$ field dependence which
essentially counts the number of vortices and a temperature dependence 
which must include a dramatic drop from $\sigma _{n}$
to $\sigma _{BS}$ as the temperature is reduced below $T_{\gamma }(H)$.
Neither the $1/H$
field dependence nor the dramatic resistivity drop below $T_{\gamma }$ is
observed in $Tl_{2}Ba_{2}CuO_{6+x}$. Further, the number of vortices cannot
differ significantly from the mean-field
estimate
because the Ginzburg parameter is small, of order $10^{-2}$ \cite{Ginzburg}.

Therefore, we conclude that the dissipartion per vortex in
$Tl_2Ba_2CuO_{6+\delta}$ is much less than that predicted by
the Bardeen-Stephen formula.
We note that an anomalously small vortex viscosity has been directly observed  
in terahertz experiments on YBCO films \cite{Parks95}.

The dissipation due to
vortex motion has been considered by many authors. A result for 2D
d-wave superconductors with circular Fermi surface and an angle independent
quasiparticle lifetime has been derived by Kopnin and Volovik \cite{Kopnin97}.
Generalizing their result to include an angular dependent lifetime, $%
\tau (\theta )$, and density of state $\nu$ yields
\begin{equation}
\sigma _{V}=\frac{1}{n_{V}}\langle \gamma (\theta )\rangle _{\theta }
\label{sigma_V}
\end{equation}
Here $\langle \rangle _{\theta }$ denotes an average over the 2D Fermi
surface parametrized by angle $\theta $ and 
\begin{equation}
\gamma (\theta )=2\nu (\theta )\Delta (\theta )^{2}\tau (\theta )\ln \left( 
\frac{\Delta _{max}^{2}}{\Delta ^{2}(\theta )}\right)   \label{gamma}
\end{equation}
Here $\Delta $ is the superconducting gap. If $\tau (\theta
)\equiv \tau $ has negligible angular dependence then $\langle \gamma
(\theta )\rangle _{\theta }$ may be reexpressed as $H_{c2}\tau $ and the
equation for $\sigma _{V}$ becomes the usual Bardeen-Stephen one. Deviations
from the Bardeen-Stephen form occur when $\tau $ has a strong angular
dependence. From Eq \ref{gamma} we see that the dissipation due to moving
vortices is determined mainly by the lifetime of the particles in the
regions where the gap is large. This will differ from the dissipation due to
the normal state conductivity, $\sigma _{n}$, if the latter is dominated by
the zone diagonals where the superconducting gap vanishes. In the usual
Boltzman approach $\sigma _{n}\approx \frac{e^{2}}{2\pi }p_{F}\langle
v_{F}(\theta )\tau (\theta )\rangle _{\theta }$;  this will be dominated by
the diagonals if
\begin{equation}
\tau (\theta )=\left( \frac{1}{\tau _{0}^{-1}+\Gamma \theta ^{2}}\right)
  \label{tau}
\end{equation}
with $\theta $ the angle measured from the zone diagonal and $\Gamma \tau
_{0}\gg 1$. This formula may be derived
by assuming that the scattering rate is the sum of two terms, a conventional
one which is due to the impurities or Fermi liquid scattering which we
parametrize by $\tau _{0}^{-1}$ and unconventional one, $\Gamma \theta ^{2}$%
, which is large everywhere except for the zone diagonals. 
Evidence for a large $\Gamma$ comes from
photoemission experiments \cite{Marshall96} which
 at $T>T_c$ find dispersing quasiparticle peaks for
momenta near the zone diagonals (``cold spots'') but find only very broad
incoherent structures for other 
momenta.  From the width of these structures we estimate $\Gamma \gtrsim
0.1\;eV$. Normal carrier conductivity implied by Eq \ref{tau}
is $\sigma _{n}=\frac{e^{2}}{\pi }v_{F}p_{F}\sqrt{\tau
_{0}/\Gamma }$. It is dominated by quasiparticles near
the zone diagonals; because the superconducting gap vanishes there 
$\sigma_n$ does not change significantly as $T$ is reduced below $T_c$.
In the $Tl_{2}Ba_{2}CuO_{6+x}$ materials of interest here
$\sigma_n$ is  of order $(10\;\mu \Omega cm)^{-1}$
implying that $\sqrt{\Gamma /\tau_0 }\approx 4 \;meV$. The 
photoemission estimate 
$\Gamma \gtrsim 0.1eV$ along with $v_F \approx1\;eV\AA$ and 
$p_F \approx 0.5\;\AA^{-1}$ 
implies $1/\tau_0 < 0.2 meV$. In a separate paper \cite
{Ioffe97} we show that Eq. (\ref{tau}) with $\tau _{0}\propto T^{2}$ and $%
\Gamma =const$ reproduces the linear resistivity and $T^{-2}$ Hall angle
observed in optimally doped materials.

Assuming $\Delta =\Delta _{0}\sin \theta $, $\nu (\theta )=\nu _{0}$
constant and $\tau (\theta )$ given by Eq (\ref{tau}) gives 
\begin{equation}
\frac{\sigma _{V}}{\sigma _{n}}=\frac{\int d\theta \sin ^{2}2\theta \tau
(\theta )}{\int d\theta \tau (\theta )}\frac{H_{c2}}{B}=\frac{\Lambda }{%
\sqrt{\Gamma \tau _{0}}}\frac{H_{c2}}{B}  \label{sigma_V/sigma_n}
\end{equation}
with $\Lambda $ a number of the order of unity determined by the detailed
behavior far from diagonals.  The
data for $Tl_{2}Ba_{2}CuO_{6+x}$ imply that $\sigma _{V}/\sigma _{n}$ is
very small for $B>0.2H_{c2}$ requiring that $\sqrt{\Gamma \tau _{0}}\gg 5$.
Our  estimates yield $\sqrt{\Gamma \tau_0} \gtrsim 20$.

Our picture may be tested in two ways. The existence of a vortex liquid in
the range $H_{\rho}<B<H_{\gamma}$ may be established by using heavy ion
irradiation to create columnar defects, pinning the vortices and raising the
melting line above $H_{\rho }(T)$.
The formula for $\sigma _{V}/\sigma _{n}$ may be tested by electron
irradiation which creates random defects which scatter electrons thereby
increasing $1/\tau _{0}$ and the normal state resistivity, thus increasing $%
\sigma _{V}/\sigma _{n}$ and hence the amplitude of magnetoresistance, etc.
Furthermore, because $\sigma _{n} \propto \tau _{0}^{1/2}$ our model predicts
that once the induced disorder becomes greater than the intrinsic one the
normal state resistivity should grow as a square root of the irradiation
time.

Now we consider the effects of inhomogeneity. We assume that the sample
contains grains of transition temperature $T_G >T_{c0}$,
size $R$ and spacing $d$, implying an
areal density $x_G=R^2/d^2$. 
The magnetization $M_{G}$ of a grain in a field $B$ is \cite{Tinkham} 
\begin{equation}
M_{G}=\frac{BR^{2}}{48\pi \lambda _{G}^{2}}  \label{M_G}
\end{equation}
with $\lambda_G$ the grain penetration depth.  The total magnetization
is then $x_GM_G$.
The experimental result is that at $T=8K
$ and $B=10\;\mbox{T}$ (much greater than the microscopic $H_{c2}$) the
diamagnetic 
magnetization $x_{G}M_{G}$ equals the London magnetization observed at the
same $T$ and $B=0.1\;\mbox{T}$. From this, Eq. (\ref{M_G}), the London formula
\cite{Tinkham} and $H_{c2}(T=8K)\approx 2\; \mbox{T}$ we conclude that 
\begin{equation}
x_{G}R^{2}=3\times 10^{4}\frac{\lambda _{G}^{2}}{\lambda ^{2}}\AA ^{2}
\end{equation}
We estimate  $\lambda _{G}/\lambda \sim 1/3 $ because 
$\lambda_G(T=0)$ should be a little
less than $\lambda(T=0)$ and the 
measurement temperature $T=8K \sim T_{c0}/2 <<T_G$; thus
$R^{2}/d\approx 100\AA $.

The above analysis assumes that the grain size is larger than the
grain coherence length $\xi_G$ which will be less than the bulk
coherence length $\xi \sim 100 \AA$
and assumes that there are no vortices in the grains.  Now in order
for a vortex to enter a grain the field must be at least large
enough that one flux quantum fits inside the grain; then there are 
additional numerical factors coming from core energy considerations
and boundary conditions. Experimentally,
the diamagnetism is linear up to at least $10;\mbox{T}$, 
thus $R^2 <\beta  \frac{\Phi_0}{10 \mbox{T}}\sim \beta%
\;20,000\;\AA^2 $ with $\beta > 1$.

The Josephson coupling between two superconducting grains in a normal metal
host depends on many details including the temperature, the magnetic field,
the size of the grain, the intergrain distance
and the strength of the electronic contact between the grain and the host
metal \cite{Tinkham}. In the dilute limit $d\gg R$ and in zero
magnetic field the Josephson energy is 
\begin{equation}
E_{J}(T,\phi )=E_{J}^{0}e^{-d/\xi _{n}}F_{d}(\phi )  \label{E_J}
\end{equation}
Here $\xi _{n}=v_{F}/(2\pi T)$ is the clean limit normal metal phase
coherence length. The phase dependence is contained in the function $%
F_{d}(\phi )$ which tends to $\sin \phi $ for $d/\xi _{n}\gg 1$ but becomes
a sawtooth function at $d/\xi _{n}\ll 1$. The Josephson energy scale $E_{J0}$
is 
\begin{equation}
E_{J0}=\Lambda ^{2}\frac{v_{F}}{2\pi d}(p_{F}R)N_{c}
\end{equation}
Here $N_{c}$ is the number of planes over which the grains extend and $%
\Lambda $ is a number which depends on the geometry and dimensions
($\Lambda \sim (R/d)^{(D-1)/2}$) and on the strength of
the electrical contact bewteen the grain and the normal metal. One expects $%
\Lambda $ to be rather less than unity.

It is convenient to define $T_{0}=v_{F}/(2\pi d)$; for $d=2000\;\AA $ and $%
R=400\;\AA $, and using $p_{F}=0.5\;\AA ^{-1}$ and $v_{F}=1\;eV\AA $ one has $%
T_{0}\approx 1K$ and $E_{J0}\approx 200\Lambda ^{2}\;K$. If the host metal
remained non-superconducting down to lowest temperatures, the grains would
order at a temperature $T_{cg}$ satisfying $T_{cg}=ZE_{J}(T_{cg})$ (here $Z$
the effective number of neighbors of a given grain,
is $\sim 6$ at $T>T_0$ and $\sim (T_0/T)^2$ for $T < T_0$). Grains in a
normal host provide an explicit realization of a superconducting
transition in which $T_c$ corresponds to a loss of phase coherence with local
pairing remaining but the temperature dependence of $\rho_S$ is due to
quasiparticles and the fluctuation regime is narrow. In the
situation of interest here, the superconductivity of the host metal is
suppressed by a magnetic field. This affects the intergrain ordering in two
ways: it frustrates the phase ordering, leading to a ``gauge glass''
behavior, and it substantially weakens the coupling between individual
grains, by causing interference between different electron paths. The gauge
glass effects on the transition temperature may be estimated by replacing $Z$
by $\sqrt{Z}$. The effect of the magnetic field
on the intergrain coupling is much larger and may be estimated by noting
that the coupling is dominated by electron trajectories in a tube of width $R
$ and length $d$; a magnetic field $B$ leads to a flux $\Phi _{B}$ through
this tube of order $\Phi _{B}=BdR$; when this flux is large compared to the
flux quantum, the usual interference arguments imply that the coupling is
reduced by the factor $\Phi _{0}/(\pi \Phi _{B})$. For $d=2000\;\AA $ and $%
R=400\;\AA $, $\Phi _{0}/(\pi dR)\approx 0.1\;\mbox{T}$, so the suppression of
the coupling in the interesting fields of order $10\;\mbox{T}$ is substantial 
(factor of 100). To summarize, the field at which the grains order is 
\begin{equation}
B_{G}(T)=\sqrt{Z}\frac{\Phi _{0}}{\pi dR}\frac{E_{J}^{0}}{T}e^{-T/T_{0}}
\label{B_G}
\end{equation}

Our numerical estimates imply that at $T=T_0\sim 1K$, $B_G \sim 20 \Lambda^2
N_c \; \mbox{T}$. In view of the large uncertainties we regard this estimate
as reasonable. The temperature dependence ($e^{-T/T_0}$ for $T>T_0$) is
certainly in qualitative agreement with the experiment.


To summarize, $Tl_{2}Ba_{2}CuO_{6+\delta }$ material exhibits two
experimental anomalies: pronounced upward curvature of $H_{c2}$ at low
temperatures and a vortex liquid regime with a negligible temperature and
field dependence of the conductivity. We argued that the former is not
intrinsic but is due to the presence in the sample of small grains with $%
T_{c}$ higher than the bulk. We estimated the density and the size of the
grains. The latter anomaly is intrinsic and important; it implies that
the vortex viscosity is unusually small. We showed that such a small
viscosity can arise if the quasiparticle relaxation rate has a very strong
anisotropy around the Fermi surface, and in particular becomes very weak
along the zone diagonal where the superconducting gap vanishes. We call the
places where the scattering rate vanishes ``cold spots''. Finally, we note
that small viscosity implies large quantum fluctuations of vortices and,
perhaps, quantum liquid of vortices at $T=0$ in a wide field range.

A crucial issue in high-$T_c$ superconductivity is the apparent
coexistence in underdoped materials of local pairing over a wide
range above the resistive $T_c$ and a mean field like superconducting
transition.  Our work shows
that there are two ways in which this may occur: the vortex
liquid contribution to the conductivity may be very small or the local pairs
can be confined to 'grains' that are coupled only via Josephson junctions
through a normal host. It will be interesting to see if these possiblities
are  indeed realized in {\it underdoped} materials.

{\bf Acknowledgements}: We thank C. Bergemann et. al. for generously sharing
their data prior to publication. A. J. M. acknowledges N.S.F. DMR-9707701;
V. G. and A. J. M. thank the physics department of Rutgers University
for hospitality.

\end{multicols}
\end{document}